\documentclass[aps,prd,preprint,superscriptaddress,amsmath,amssymb,showpacs]{revtex4-1}
\usepackage{dcolumn}
\usepackage{graphicx}
\usepackage{float}
\usepackage{physics}
\usepackage{xcolor}
\usepackage{multirow}
\usepackage{lastpage}
\usepackage{longtable}
\usepackage{amsmath}
\usepackage[colorlinks=true,allcolors=blue]{hyperref}

\newcommand{\orcid}[1]{\href{https://orcid.org/#1}{\includegraphics[scale=0.055]{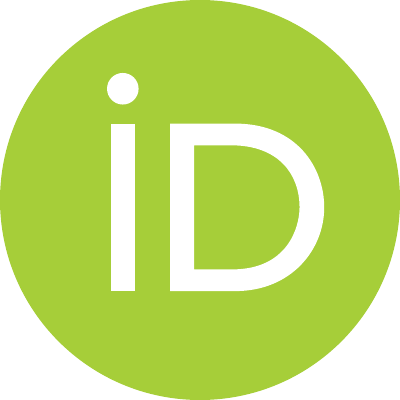}}}

\graphicspath{{figs/}}
\begin{document}
\title{Bayesian neural network with autoencoder for model-based description of $\alpha$-particle preformation factor}
\author{Xiao-Yan Zhu\hspace*{0.6pt}\orcid{0009-0008-2830-9866}\hspace*{0.8pt}}
\email[Electronic address: ]{xyzhu0128@163.com}
\affiliation{School of Mathematics and Physics, University of South China, Hengyang, 421001, People's Republic of China}
\author{Heng-Jian Si-Tu}
\affiliation{School of Mathematics and Physics, University of South China, Hengyang, 421001, People's Republic of China}
\author{Hao Zhang}
\affiliation{School of Nuclear Science and Technology, University of South China, Hengyang, 421001, People's Republic of China}
\author{Wei Gao\hspace*{0.6pt}\orcid{0009-0007-8823-5240}\hspace*{0.8pt}}
\email[Electronic address: ]{weigao@my.swjtu.edu.cn}
\affiliation{School of Physical Science and Technology, Southwest Jiaotong University, Chengdu, 610031, People's Republic of China}
\author{Wen-Bin Lin\hspace*{0.6pt}\orcid{0000-0002-4282-066X}\hspace*{0.8pt}}
\email[Electronic address: ]{lwb@usc.edu.cn}
\affiliation{School of Mathematics and Physics, University of South China, Hengyang, 421001, People's Republic of China}
\author{Xiao-Hua Li\hspace*{0.6pt}\orcid{0000-0002-3399-6057}\hspace*{0.8pt}}
\email[Electronic address: ]{lixiaohuaphysics@126.com}
\affiliation{School of Nuclear Science and Technology, University of South China, Hengyang, 421001, People's Republic of China}

\date{\today}
\begin{abstract}
$\alpha$ decay is an important probe for studying the structure of heavy and superheavy nuclei, in which the $\alpha$-particle preformation ($P_{\alpha}$) is a key physical quantity for describing decay half-lives. This work develops a hybrid framework that integrates Bayesian neural networks with autoencoder (BNN-Auto), combined with the cosh potential (CPT), to systematically optimize the constraint and prediction of $P_{\alpha}$. The model employs variational inference for probabilistic modeling of network weights, naturally providing robust uncertainty quantification for predictions, and utilizes an autoencoder to enhance the robustness of feature representation. Based on experimental data from 535 nuclei, the BNN-Auto method achieves relative improvements in the root mean square deviation ($\sigma_{\rm{RMS}}$) of $P_{\alpha}$ prediction by $61.14\%$ on the training set and $54.49\%$ on the validation set. Further analysis reveals that the $P_{\alpha}$ and half-life extracted by the model exhibit pronounced odd-even staggering and shell effects in isotopic chains with $Z=86-90$ and isotones with $N=124-128$ and $N=150-154$. Moreover, we successfully predict the $\alpha$ decay half-lives of nuclei with $Z=120$ and observe a significant increase in the half-life near $N=184$, which verifies the shell effect of the predicted 'stable island'. This study not only provides a high-precision theoretical description for $\alpha$ decay, but also offers a new machine learning perspective for exploring the structure of superheavy nuclei.
\end{abstract}

\maketitle
\section{Introduction}
Since the successful explanation of $\alpha$ decay by Gamow \cite{Gamow:1928zz}, Condon and Gurney \cite{Gurney:1928lxa} using quantum tunneling effect, it has remained a pivotal physical process in the study of heavy and superheavy nuclei \cite{Carroll:2014yga,Delion:2015mnw,Seif:2011zz,Ren:2012zza}. $\alpha$ decay serves not only as an indispensable experimental tool for identifying newly synthesized nuclides, especially superheavy elements with $Z>118$, but also as a sensitive probe for testing and refining nuclear structure models \cite{Hofmann:2000cs,Andreyev:2013iwa,ahmed2013clusterization,Ni:2015aha}. 
Investigations into this process help extract important structural information such as nuclear shell effects and deformation \cite{Bao:2017gln}. Theoretical developments in this field have primarily focused on core issues including the calculation of decay energies, the determination of half-lives, and the $\alpha$-particle preformation probability \cite{Zhang:2008ct,Dong:2010pw,Guo:2014era,Bao:2014wxa,wang2015systematic}. 

In theoretical descriptions, the half-lives of $\alpha$ decay is commonly expressed as the product of two core physical quantities: the preformation probability $P_{\alpha}$ of the $\alpha$ particle and its probability $P$ of penetrating the potential barrier. The penetration probability $P$ depends primarily on the decay energy and the shape of the barrier and can treated with relatively unified methods such as the WKB approximation. In contrast, the preformation probability $P_{\alpha}$, involves the degree of $\alpha$-clustering at the surface of the parent nucleus, is a complex quantum many-body problem closely linked to microscopic nuclear structure information such as nucleon correlations and shell effects. It is also the key point of divergence among different theoretical models. Within this framework, a variety of macroscopic-microscopic models have been developed, including the double-folding potential method \cite{Zhang:2009aq,Ghodsi:2021gwt,NhuLe:2025okf,Zheng:2025yac}, the density-dependent cluster model \cite{Xu:2005hlv,Tonozuka:1979esf,Kobos:1984zz,Zhang:2013xwa}, and the generalized liquid-drop model \cite{Royer:2000zi,Zhang:2006dj,Bao:2012hgp,Santhosh:2019plu}. These models differ significantly in their descriptions of $P_{\alpha}$, typically relying on specific nuclear structure assumptions or empirical parameterizations \cite{Varga:1992oju,Delion:2004wv,Delion:2007zz,lovas1998microscopic,Santhosh:2014zha}. As a result, the extracted preformation factors exhibit marked model dependence. Therefore, accurate theoretical description of $\alpha$ decay half-lives hinges essentially on the ability of a model to effectively capture the microscopic nuclear structure information associated with $P_{\alpha}$.

However, direct calculation of $P_{\alpha}$ in microscopic theory is extremely complex and relies heavily on model assumption. Therefore, in extensive phenomenological analysis and predictive applications, an indirect approach is commonly adopted, that is, by inversely deducing the ratio of the theoretically calculated $\alpha$ decay half-life to the experimental one. In this processing, $P_{\alpha}$ is regarded as an adjustable factor related to the nuclear structure, and is even approximated as a constant in different models \cite{Zhang:2011aq,Tang:2023nap,SalehAhmed:2017qzf,Deng:2015qha}. Nevertheless, the reliability of extrapolating $P_{\alpha}$ and half-life for unknown nuclei using this method depends critically on the selected systematic laws, and its generalizability is particularly limited in the region of superheavy nuclei far from the stability line. Recently, a preformation factor model based on the least squares method combined with the cosh model in the two-potential (CPT-LSM) approach has been proposed to systematically describe $P_{\alpha}$ \cite{Luo:2024ogt}, yet it cannot be extended globally to all nuclei, which reflects the limitations of traditional parameterized or semi-empirical theoretical models. Thus, there is an urgent need to establish a novel method capable of adaptively learning the systematic trends of nuclear structure from data, in order to globally describe $P_{\alpha}$ and accurately predict the $\alpha$ decay half-lives of superheavy nuclei.

Among various machine learning methodologies \cite{Carleo:2019ptp,Pang:2016vdc,Krastev:2019koe,Soma:2022vbb,Lasseri:2019ywk,Negoita:2018kgi,wang:2019pct,
Niu:2018trk,Niu:2018olp,Ma:2019bbf,Ma:2023ofi,Luo:2025dki,Zhu:2025ujz}, the Bayesian neural network (BNN) is distinguished by its inherent probabilistic framework, which treats network weights as random variables and performs modeling through Bayesian inference. This approach naturally provides uncertainty estimates for predictions, mitigates overfitting via automatic complexity control, and enhances generalization capability \cite{Utama:2015hva,Utama:2016tcl,Neufcourt:2018syo,Neufcourt:2019qvd,Dong:2021aqg,Liu:2025gdl,Jin:2023igd}.  These characteristics enable them to robustly extract key information on underlying physical laws from large datasets. In contrast, traditional neural networks rely on fixed weights, lack uncertainty quantification, and tend to overfit, especially in limited data such as studies of the $\alpha$-particle preformation mechanism and half-life calculations \cite{Liu:2025gdl,Luo:2025dki}. To address these limitations and improve the reliability of half-half predictions, this study develops a BNN model integrated with an autoencoder architecture (BNN-Auto), based on the physical description of $\alpha$ decay, to systematically constrain and predict the $P_{\alpha}$, thereby achieving precise calculations of $\alpha$ decay half-lives. Furthermore, it facilitates the exploration of nuclear structure information embedded within the model such as the evident odd-even staggering effects and shell effects observed in isotopic chains with proton numbers $Z=86-90$ and in isotones with neutron numbers $N=124-128$  and $N=150-154$. Thus, this paper aims to evaluate the effectiveness and applicability of such hybrid neural network frameworks in investigating the $P_{\alpha}$.

The structure of this article is organized as follows. Section \ref{Sec.II} gives a brief introduction of the theoretical framework, including the $\alpha$ decay model with the cosh potential and the Bayesian neural network with autoencoder. In Section \ref{Sec.III}, the detailed calculations and discussions are presented. Finally, Section \ref{Sec.IV} provides a concise summary.

\section{Theoretical framework}\label{Sec.II}
\subsection{$\alpha$ decay model with a cosh potential }\label{Sec.II.1}
In the process of $\alpha$ decay, the total interaction potential between the $\alpha$ particle and daughter nucleus is the sum of the nuclear, Coulomb, and centrifugal potential barrier. It can be written as
\begin{equation}
 V(r)=V_N(r)+V_C(r)+V_l(r),\label{V}
\end{equation}
where $V_N(r)$ denotes the nuclear potential. In this work, the cosh potential (CPT) is adopted for $V_N(r)$, expressed as \cite{Buck:1992zz}
\begin{equation}
 V_N(r)=-V_0\frac{1+\mathrm{cosh}(R/a_0)}{\mathrm{cosh}(r/a_0)+\mathrm{cosh}(R/a_0)},\label{Vn}
\end{equation}
where the depth $V_0$ and diffuseness $a_0$ of the nuclear potential are parametrized as $a_0=0.5958$ fm and $V_0=192.42+31.059(N_d-Z_d)/A_d$ MeV \cite{Sun:2015wdu}, with $N_d$, $Z_d$ and $A_d$ denoting the neutron, proton, and mass numbers of the daughter nucleus, respectively. The term $V_C(r)$ represents the Coulomb potential of a uniformly charged sphere with a sharp radius $R$, and is given by
\begin{eqnarray}
V_c(r) & = & \left\{\begin{array}{ll}
\frac{Z_{\alpha}Z_d e^2}{2R}\big(3-\frac{r^2}{R^2}\big), & r\leq R, \\
\frac{Z_{\alpha}Z_d e^2}{r}, & r\textgreater R,
\end{array}\right.
\end{eqnarray}
where $Z_{\alpha}$ denotes the proton number of $\alpha$-particle, the radius parameter is given by $R=1.28A^{1/3}-0.76+0.8A^{-1/3}$ fm. The centrifugal potential is treated in the Langer approximation, $V_l(r)=\frac{\hbar^2(l+\frac{1}{2})^2}{2\mu r^2}$, with $l$ being the orbital angular momentum.

For the decay of a quasistationary state lying between the bound and scattering regimes, as treated by Gurvitz and Kalbermann \cite{Gurvitz:1986uv}, the Gamow formula is modified by introducing a well-defined pre-exponential factor. The decay width $\Gamma$ is then obtained as
\begin{equation}
   \Gamma=\frac{\hbar^2F }{4\,\mu}\,P_{\alpha}\mathrm{exp}\bigg[-2\int_{r_2}^{r_3}k(r)\,dr\bigg] ,\label{Gamma}
\end{equation}
where $\mu$ is the reduced mass of the $\alpha$-particle and daughter nucleus, and $F$ is the normalization factor, defined as 
\begin{equation}
    F\int_{r_1}^{r_2}\frac{1}{2\,k(r)}dr=1,\label{F}
\end{equation}
where $k(r)=\sqrt{\frac{2\mu}{\hbar^2}|V(r)-Q_{\alpha}|}$ represents the wave number. The three classical turning points $r_1$, $r_2$ and $r_3$ are determined by the condition $V(r)=Q_{\alpha}$, with $Q_{\alpha}$ being the $\alpha$ decay energy. The corresponding half-life is then $T_{1/2}=\hbar \,\mathrm{ln2}/\Gamma$.

The microscopic calculation of $P_{\alpha}$ is highly complex, and its theoretical value is often difficult to obtain directly. To extract this key physical quantity from experimental data, a commonly employed semi-empirical approach is to derive it based on $\alpha$ decay half-lives. Specifically, in theoretical calculations, it is assumed that $P_{\alpha}^{\mathrm{the}}=1$, and the corresponding calculated half-life denoted as $T_{1/2}^{\mathrm{cal}}$. However, the experimental half-life $T_{1/2}^{\mathrm{exp}}$ is usually shorter, indicating that the actual preformation probability is less than 1. Therefore, the experimental preformation factor $P_{\alpha}^{\mathrm{exp}}$ can be described by 
\begin{equation}
 P_{\alpha}^{\mathrm{exp}}=\frac{T_{1/2}^{\mathrm{cal}}}{T_{1/2}^{\mathrm{exp}}},\label{Paexp}
\end{equation}

\subsection{Bayesian neural network with autoencoder }\label{Sec.II.2}
BNN constitutes a probabilistic modeling framework grounded in the principles of Bayesian statistics \cite{neal:1996bayesian}. Central to this overview is Bayes' theorem, which serves as the foundation for deriving the posterior distribution $p(\omega|\mathcal{D})$ over network weights $\omega$ given the dataset $\mathcal{D}$. It is written by
\begin{equation}
    p(\omega|\mathcal{D})=\frac{p(\mathcal{D}|\omega)p(\omega)}{p(\mathcal{D})},\label{Bayes}
\end{equation}
where $p(\omega)$ represents the prior distribution over the weights, $p(\mathcal{D}|\omega)$ is the likelihood of the data, and $p(\mathcal{D})$ is the marginal likelihood. Within this framework, a zero-mean Gaussian prior distribution $p(\omega)$ is typically placed on the network weights. For a regression task, the likelihood is often assumed to be Gaussian: $p(\mathcal{D}|\omega) = \prod_{i} \mathcal{N}(y_i | H(\mathbf{x}_i, \omega), \sigma^2)$, where $(\mathbf{x}_i, y_i)$ is a data point, $H$ is the neural network function, and $\sigma^2$ is the noise variance.

Given that the exact posterior is computationally intractable, we employ variational inference. A variational distribution $q(\omega|\theta)$ parameterized by $\theta$ is introduced, and $\theta$ is optimized by minimizing the Kullback-Leibler (KL) divergence to the true posterior:
\begin{equation}
    \mathrm{KL}(q(\omega|\theta)\|p(\omega|\mathcal{D}))=\int q(\omega|\theta) \log\frac{q(\omega|\theta)}{p(\omega|\mathcal{D})} d\omega.\label{KL}
\end{equation}
The prediction for a new input $\mathbf{x}^*$ is given by the posterior predictive distribution, which marginalizes over the uncertain parameters:
\begin{equation}
    p(y^* | \mathbf{x}^*, \mathcal{D}) = \int p(y^* | \mathbf{x}^*, \omega) p(\omega|\mathcal{D}) d\omega \approx \frac{1}{T} \sum_{t=1}^{T} H(\mathbf{x}^*, \omega^{(t)}), \label{predictive}
\end{equation}
where $\omega^{(t)} \sim q(\omega|\theta)$ represents the network parameters sampled from the posterior distribution, and the integral is approximated via Monte Carlo sampling.

\begin{figure}[!htb]\centering
 \includegraphics
  [width=1.0\hsize]
  {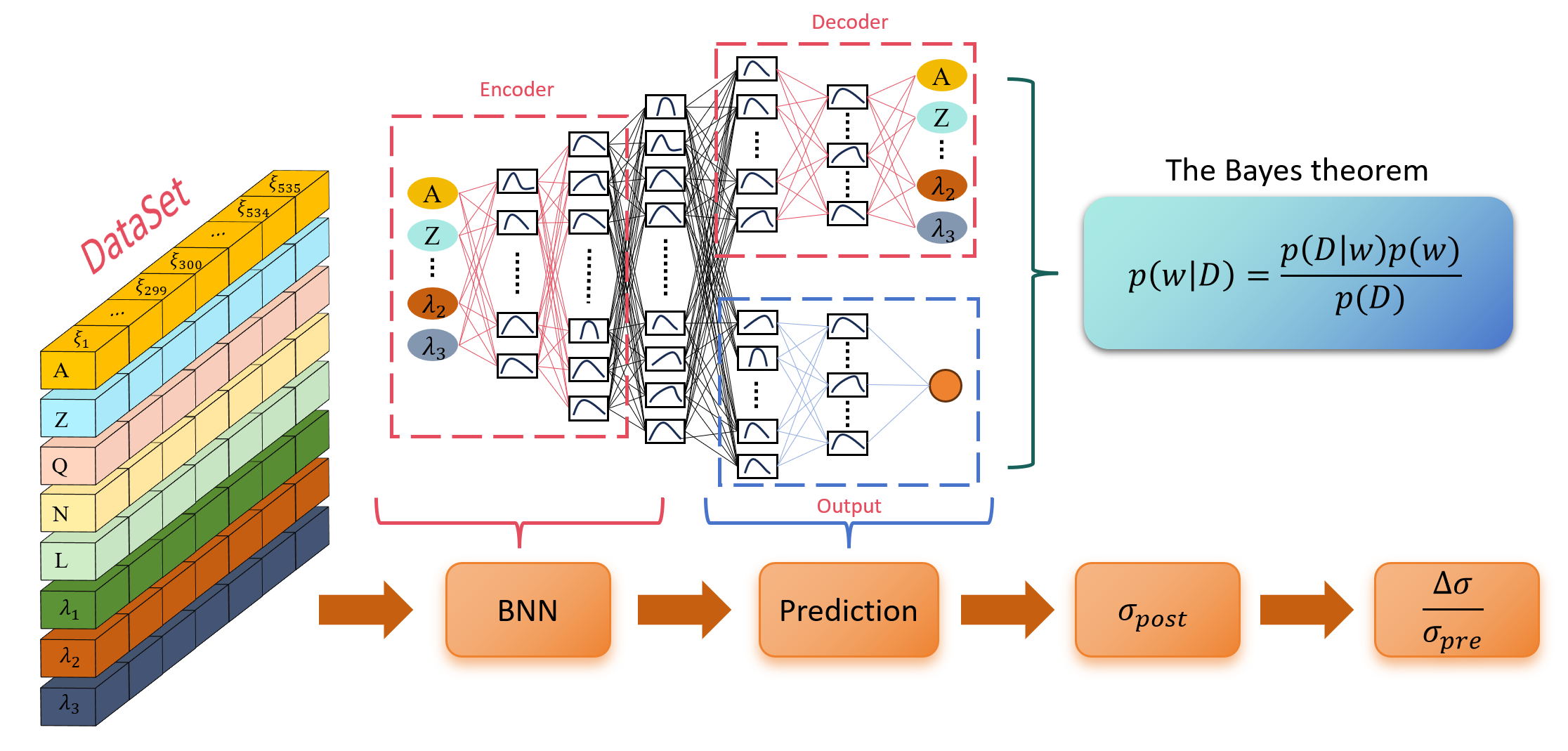}
 \caption{The flowchart of the BNN-autoencoder structure based on Bayes' theorem, from data encoding to probability prediction}
 \label{Fig-1}
 \end{figure}

In the present work, the input feature vector $\mathbf{x}$ for a nucleus is constructed to encapsulate key nuclear properties. It comprises eight primary features: the mass number $A$, neutron number $N$, and proton number $Z$ of the parent nucleus, $Q_{\alpha}$, $l$, and three derived features, namely $\lambda_1=A^{1/3}$, $\lambda_2=Z/\sqrt{Q_{\alpha}}$, $\lambda_3=\sqrt{l(l+1)}$. These features correspond to the main physical factors governing $P_{\alpha}$. As illustrated in Fig. \ref{Fig-1}, the network consists of three core components: an encoder, a decoder, and a prediction head. The encoder is composed of three fully connected layers, each followed by batch normalization and a ReLU activation function. The decoder symmetrically reconstructs the input by employing a Sigmoid activation function. The prediction head utilizes an ELU activation function to map the high-dimensional latent variables to the final predicted $P_{\alpha}$.

During the conversion of the deterministic network into a BNN, a probability distribution is placed over all weight parameters via the reparameterization trick. The variational parameters are optimized by minimizing the loss function $L$, which integrates a prediction error term ($L_{\mathrm{pre}}$), a reconstruction error from the autoencoder ($L_{\mathrm{recon}}$), and the KL divergence regularization term:
\begin{equation}
L = k L_{\mathrm{recon}} + (1-k) L_{\mathrm{pre}} + \mathrm{KL}(q(\omega|\theta)\|p(\omega)).\label{L}
\end{equation}
Here, $k$ denotes a weighting factor that balances the reconstruction and prediction losses. The prediction loss $L_{\mathrm{pre}}$ typically corresponds to the negative log-likelihood. The model is trained using the AdamW optimizer with a learning rate of $10^{-4}$, weight decay of $10^{-6}$, and a batch size of 32. To evaluate stability, 10-fold cross-validation repeated 5 times is performed, with training conducted over 1000 epochs \cite{adam:2014method,tian:2022comprehensive}.

\section{Results and discussion}\label{Sec.III}
In this work, it is achieved a high-precision description of the $\alpha$-particle preformation factor in $\alpha$ decay half-lives by integrating a hybrid framework combining Bayesian neural networks with autoencoders (BNN-Auto) based on the cosh potential (CPT) model. The $\alpha$ preformation factor is a key physical quantity in nuclear decay theory, directly linked to nuclear structure information such as shell effects and nucleon correlations. However, traditional models exhibit limitations due to the complexity and model dependency \cite{Deng:2020rzy,Deng:2021siq,Luo:2024ogt}. The BNN–Auto approach provides predictive uncertainty quantification within a Bayesian probabilistic framework, while the autoencoder enhances the robustness of feature representation, thereby uncovering hidden physical patterns in a data-driven manner. This work begins by extracting experimental preformation factors $P_{\alpha}^{\mathrm{exp}}$ from the experimental half-lives data of 535 nuclei, including 159 even–even, 295 odd-A, and 81 odd–odd nuclei using Eq. (\ref{Paexp}). Input features comprise physical quantities such as proton number $N$, neutron number $Z$, $\alpha$ decay energy $Q_{\alpha}$, and angular momentum $l$, which are fed into the BNN–Auto model for training and prediction. The network architecture, illustrated in Fig. \ref{Fig-1}, employs an encoder–decoder structure to effectively capture nonlinear relationships among nuclear features, whereas the Bayesian framework supplies posterior distributions for predictions by variational inference. To evaluate the performance of the model, the root mean square deviation $\sigma_{\rm{RMS}}$ is adopted as the indicator, which is defined as
\begin{equation}
\sigma_{\rm{RMS}}=\sqrt{\frac{1}{n}\sum(\mathrm{log_{10}}P_{\alpha}^{\mathrm{BNN-Auto}}-\mathrm{log_{10}}P_{\alpha}^{\mathrm{exp}})^2}.\label{sigma}
\end{equation}
Here, $\mathrm{log_{10}}P_{\alpha}^{\mathrm{BNN-Auto}}$ is the logarithm of preformation factor obtained by BNN-Auto method. During the prediction phase, the mean and standard deviation are calculated through 20 stochastic forward passes to quantify the uncertainty:
$P_{\alpha}=\sum_{t=1}^{20}H(x^*;\omega^{(t)})/20$, $\sigma=\sum_{t=1}^{20}(H(x^*;\omega^{(t)})-P_{\alpha})^2/19$. The relative improvement ratio, defined as $\frac{\Delta\sigma} {\sigma_{\mathrm{pre}}}=\frac{\sigma_{\mathrm{pre}}-\sigma_{\mathrm{post}}}{\sigma_{\mathrm{pre}}}$, is used to assess the enhancement offered by the BNN–Auto method over traditional approaches such as the LSM combined with the CPT model (CPT–LSM) \cite{Luo:2024ogt}. Here $\sigma_{\mathrm{pre}}$ denotes the RMS between the $\alpha$-particle preformation factors calculated using the cosh-potential model with LSM fitting and the experimental ones, whereas $\sigma_{\mathrm{post}}$ represents the corresponding RMS obtained with the BNN–Auto method. As shown in the comparative results in Table \ref{table1}, the BNN–Auto method significantly improves predictive consistency. For even–even nuclei, the relative improvement reaches $61.14\%$ on the training set and $54.49\%$ on the validation set. In the case of odd-A nuclei, the improvements are $57.96\%$ and $35.48\%$, respectively, while for odd–odd nuclei, they reach $56.80\%$ and $66.74\%$. Overall, across all nuclei, the method yields an improvement of $30.75\%$ on the training set and $35.37\%$ on the validation set. These results demonstrate that the BNN–Auto framework not only possesses a strong global optimization capability but also performs robustly in extrapolative testing. It effectively captures local interdependencies in the preformation factors among different nuclides, thereby refining the outcomes of physical calculations.

\begin{table}[h]
	\centering
	\setlength{\tabcolsep}{12pt}
	\caption{The RMS deviation $\sigma_{\rm{pre}}$ of $\rm{log_{10}}P_{\alpha}$ calculated by the CPT–LSM \cite{Luo:2024ogt} and $\sigma_{\text{post}}$ after BNN-Auto corrections for 535 nuclei.}
	\begin{tabular}{ccccccc}
			\toprule
			\multirow{2}{*}{\textbf{Modes}} & \multicolumn{3}{c}{\textbf{Learning Set}} & \multicolumn{3}{c}{\textbf{Validation Set}} \\
			\cline{2-7}
			& $\sigma_{\text{pre}}$ & $\sigma_{\text{post}}$ & $\frac{\Delta\sigma}{\sigma_{\text{pre}}}$ &    $\sigma_{\text{pre}}$ & $\sigma_{\text{post}}$ & $\frac{\Delta\sigma}{\sigma_{\text{pre}}}$ \\
			\hline
			even-even nuclei & 0.404 & 0.157 & $61.14\%$ & 0.345 & 0.157 & $54.49\%$ \\
			odd-A nuclei& 0.471 & 0.198 &$57.96\%$  &0.372 & 0.240 &$35.48\%$  \\
			odd-odd nuclei& 0.588 & 0.254 & $56.80\%$ & 0.427 & 0.142 & $66.74\%$ \\
			all nuclei& 0.465 & 0.322 & $30.75\%$ &0.458  & 0.296 & $35.37\%$ \\\hline
	\end{tabular}
	\label{table1}
\end{table}

\begin{figure}[!htb]\centering
 \includegraphics
  [width=0.85\hsize]
  {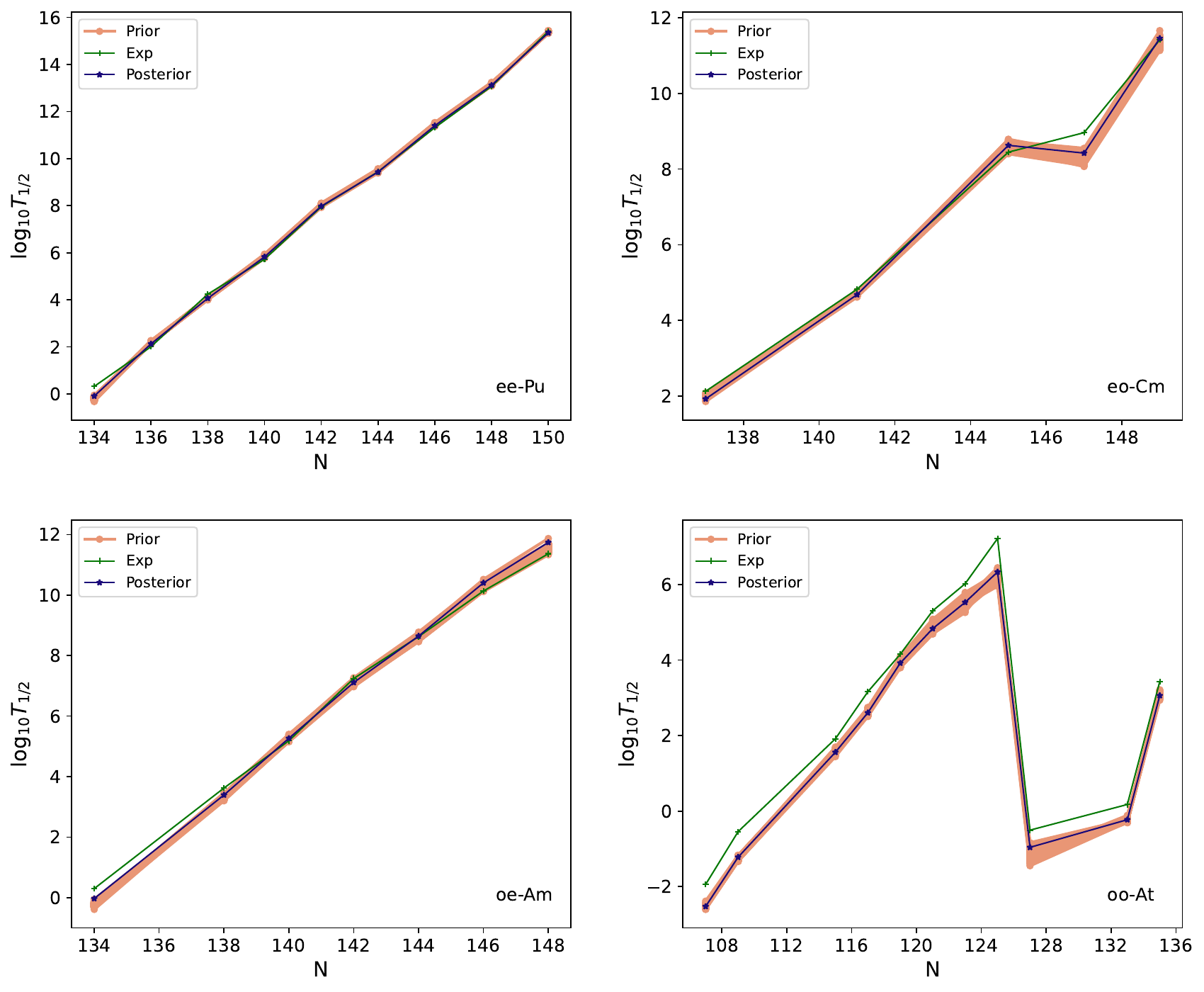}
 \caption{Prior and posterior distributions of $\rm{log_{10}}{T}_{1/2}^{\rm{BNN-Auto}}$ for even-even Pu, odd-A Cm and Am, and odd-odd At isotopes.}
 \label{Figure2}
 \end{figure}

The above results indicate that BNN-Auto significantly enhances the accuracy of predictions and offers new insights into the physical mechanism of the $\alpha$ preformation factor. Although current theoretical models capture the essential physical characteristics of $\alpha$ decay, there are still exhibit certain deviations in calculating half-lives, which primarily stem from the influence of nuclear shell structure effects. To more comprehensively incorporate such nuclear structure effects, it is necessary to refine existing $\alpha$ decay theoretical models using the BNN-Auto. Based on the $P_{\alpha}^{\mathrm{BNN-Auto}}$ obtained from BNN–Auto, the $\alpha$ decay half-lives $\rm{log}_{10}{T}_{1/2}^{\rm{BNN-Auto}}$ are further calculated. As shown in Table \ref{table2}, taking the even-even (ee) Pu of $Z=94$, odd-A including even-odd (eo) Cm of $Z=96$, odd-even (oe) Am of $Z=95$, as well as the odd-odd (oo) At of $Z=85$ nuclei as examples, the modified half-lives are in excellent agreement with the experimental data. Simultaneously, Fig. \ref{Figure2} shows the comparison of the calculated distribution of the logarithmic half-lives for ee-Pu, eo-Cm, oe-Am, and oo-At isotopes. In this figure, the initial theoretical calculations (prior) based on the CPT model are shown in orange, the experimental ones are plotted in green as the benchmark, and the posterior distributions corrected by the BNN-Auto method are depicted in dark blue. It can be clearly observed that for all four types of nuclei, the orange color representing the prior theoretical calculation usually deviates significantly from the green experimental, and the implied prediction uncertainty range is relatively wide. In contrast, the dark blue posterior calculations obtained after training within the BNN-Auto framework shift significantly toward the experimental green line, nearly overlapping with it. Moreover, the posterior uncertainty is substantially narrower compared to the prior. This intuitively proves that the BNN-Auto method can effectively integrate experimental data, precisely calibrate the theoretical model, and significantly improve the prediction accuracy and reliability of the half-life of $\alpha$ decay for various nuclides, including complex odd-A nuclides and odd-odd nuclides.

\renewcommand\arraystretch{0.5}
\setlength{\tabcolsep}{3.6mm}
\begin{longtable}{ccccccc}
	\caption{Calculations of $\alpha$-particle preformation factors and half-lives for Pu, Cm, Am, and At within BNN-Auto method. The experimental $\alpha$ decay energies and half-lives are taken from Refs. \cite{Deng:2020rzy,Luo:2025dki,Huang:2021nwk,Wang:2021xhn,Zhu:2024swx}}\\
	\hline
    \toprule
	{$\alpha$ transition}  & $Q_{\alpha}$  & $l$ & $\rm{log}_{10}{T}_{1/2}^{\rm{exp}}$ &$P_{\alpha}^{\rm{exp}}$&$P_{\alpha}^{\rm{BNN-Auto}}$ & $\rm{log}_{10}{T}_{1/2}^{\rm{BNN-Auto}}$\\
 	\hline
	\endfirsthead
	\multicolumn{1}{l}
	{Continued.} \\
\hline
    \toprule
	{$\alpha$ transition}  & $Q_{\alpha}$  & $l$ & $\rm{log}_{10}{T}_{1/2}^{\rm{exp}}$ &$P_{\alpha}^{\rm{exp}}$&$P_{\alpha}^{\rm{BNN-Auto}}$ & $\rm{log}_{10}{T}_{1/2}^{\rm{BNN-Auto}}$\\
	\hline
	\endhead
	\hline \\
	\endfoot
	\endlastfoot
$	^{	228	}	$Pu$	\to	^{	224	}	$U$	$	&	7.9403	&	0	&	0.3222 	&	0.0773 	&	0.1989 	&	-0.0885 	\\
$	^{	230	}	$Pu$	\to	^{	226	}	$U$	$	&	7.1803	&	0	&	2.0086 	&	0.8088 	&	0.6132 	&	2.1289 	\\
$	^{	232	}	$Pu$	\to	^{	228	}	$U$	$	&	6.7163	&	0	&	4.2413 	&	0.3373 	&	0.5065 	&	4.0648 	\\
$	^{	234	}	$Pu$	\to	^{	230	}	$U$	$	&	6.3103	&	0	&	5.7226 	&	0.7186 	&	0.5669 	&	5.8256 	\\
$	^{	236	}	$Pu$	\to	^{	232	}	$U$	$	&	5.8674	&	0	&	7.9552 	&	0.6908 	&	0.6541 	&	7.9788 	\\
$	^{	238	}	$Pu$	\to	^{	234	}	$U$	$	&	5.5935	&	0	&	9.4421 	&	0.6698 	&	0.6909 	&	9.4286 	\\
$	^{	240	}	$Pu$	\to	^{	236	}	$U$	$	&	5.2561	&	0	&	11.3161 	&	0.8696 	&	0.7188 	&	11.3988 	\\
$	^{	242	}	$Pu$	\to	^{	238	}	$U$	$	&	4.9844	&	0	&	13.0731 	&	0.8427 	&	0.7602 	&	13.1179 	\\
$	^{	244	}	$Pu$	\to	^{	240	}	$U$	$	&	4.6658	&	0	&	15.4022 	&	0.7231 	&	0.8280 	&	15.3434 	\\
$	^{	233	}	$Cm$	\to	^{	229	}	$Pu$	$	&	7.4653 	&	0	&	2.1282 	&	0.3479 	&	0.5541 	&	1.9261 	\\
$	^{	237	}	$Cm$	\to	^{	233	}	$Pu$	$	&	6.7753 	&	0	&	4.8239 	&	0.3347 	&	0.4650 	&	4.6811 	\\
$	^{	241	}	$Cm$	\to	^{	237	}	$Pu$	$	&	6.1854 	&	3	&	8.4481 	&	0.1164 	&	0.0769 	&	8.6281 	\\
$	^{	243	}	$Cm$	\to	^{	239	}	$Pu$	$	&	6.1690 	&	2	&	8.9630 	&	0.0226 	&	0.0786 	&	8.4227 	\\
$	^{	245	}	$Cm$	\to	^{	241	}	$Pu$	$	&	5.6247 	&	2	&	11.4155 	&	0.0644 	&	0.0587 	&	11.4556 	\\
$	^{	229	}	$Am$	\to	^{	225	}	$Np$	$	&	8.1353 	&	2	&	0.2962 	&	0.0806 	&	0.1717 	&	-0.0325 	\\
$	^{	233	}	$Am$	\to	^{	229	}	$Np$	$	&	7.0553 	&	1	&	3.6215 	&	0.1832 	&	0.3115 	&	3.3910 	\\
$	^{	235	}	$Am$	\to	^{	231	}	$Np$	$	&	6.5853 	&	1	&	5.1835 	&	0.4802 	&	0.3988 	&	5.2642 	\\
$	^{	237	}	$Am$	\to	^{	233	}	$Np$	$	&	6.1953 	&	1	&	7.2419 	&	0.2533 	&	0.3430 	&	7.1103 	\\
$	^{	239	}	$Am$	\to	^{	235	}	$Np$	$	&	5.9226 	&	1	&	8.6275 	&	0.2407 	&	0.2317 	&	8.6441 	\\
$	^{	241	}	$Am$	\to	^{	237	}	$Np$	$	&	5.6380 	&	1	&	10.1352 	&	0.2545 	&	0.1370 	&	10.4040 	\\
$	^{	243	}	$Am$	\to	^{	239	}	$Np$	$	&	5.4392 	&	1	&	11.3662 	&	0.2056 	&	0.0868 	&	11.7408 	\\
$	^{	192	}	$At$	\to	^{	188	}	$Bi$	$	&	7.6963 	&	0	&	-1.9393 	&	0.0967 	&	0.3714 	&	-2.5237 	\\
$	^{	194	}	$At$	\to	^{	190	}	$Bi$^{\rm{n}}$$	$	&	7.3343 	&	0	&	-0.5436 	&	0.0560 	&	0.2655 	&	-1.2194 	\\
$	^{	200	}	$At$	\to	^{	196	}	$Bi$	$	&	6.5963 	&	0	&	1.9170 	&	0.0697 	&	0.1584 	&	1.5606 	\\
$	^{	202	}	$At$	\to	^{	198	}	$Bi$	$	&	6.3533 	&	0	&	3.1610 	&	0.0358 	&	0.1284 	&	2.6060 	\\
$	^{	204	}	$At$	\to	^{	200	}	$Bi$	$	&	6.0713 	&	0	&	4.1561 	&	0.0566 	&	0.0971 	&	3.9217 	\\
$	^{	206	}	$At$	\to	^{	202	}	$Bi$	$	&	5.8863 	&	0	&	5.3058 	&	0.0280 	&	0.0830 	&	4.8342 	\\
$	^{	208	}	$At$	\to	^{	204	}	$Bi$	$	&	5.7513 	&	0	&	6.0234 	&	0.0214 	&	0.0666 	&	5.5300 	\\
$	^{	210	}	$At$	\to	^{	206	}	$Bi$	$	&	5.6313 	&	2	&	7.2213 	&	0.0091 	&	0.0704 	&	6.3342 	\\
$	^{	212	}	$At$	\to	^{	208	}	$Bi$	$	&	7.8173 	&	5	&	-0.5031 	&	0.0091 	&	0.0260 	&	-0.9589 	\\
$	^{	218	}	$At$	\to	^{	214	}	$Bi$	$	&	6.8743 	&	0	&	0.1761 	&	0.1583 	&	0.4000 	&	-0.2264 	\\
$	^{	220	}	$At$	\to	^{	216	}	$Bi$^{\rm{m}}$$	$	&	6.0533 	&	0	&	3.4337 	&	0.1742 	&	0.4104 	&	3.0615 	\\
\toprule	
\label{table2}
\end{longtable}

\begin{figure}[!htb]\centering
 \includegraphics
  [width=0.9\hsize]
  {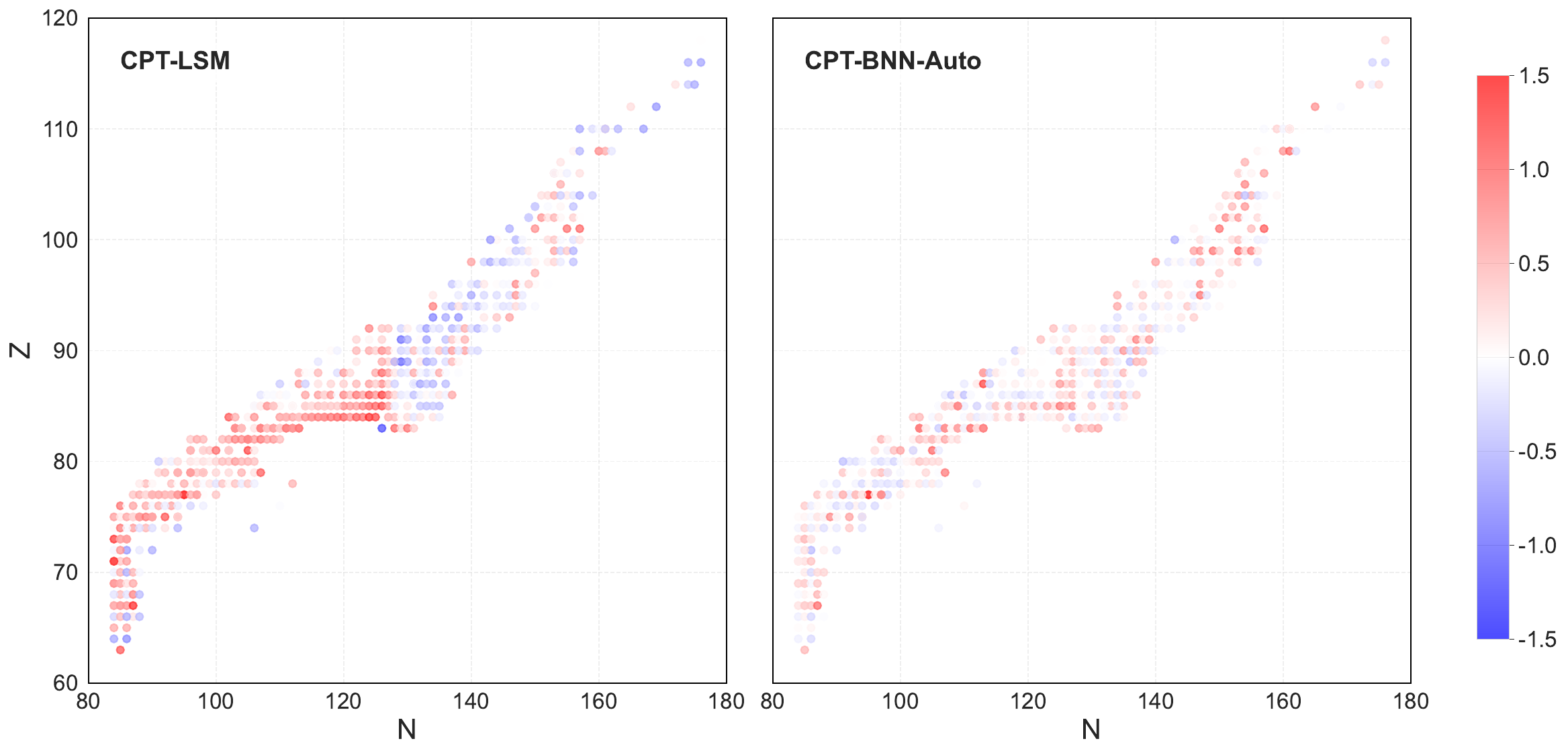}
 \caption{Residuals of the logarithmic $\alpha$ decay half-lives, differences between experimental ones and theoretical predictions from the CPT-LSM and CPT-BNN-Auto methods.}
 \label{Figure3}
 \end{figure}

\begin{figure}[!htb]\centering
 \includegraphics
  [width=0.6\hsize]
  {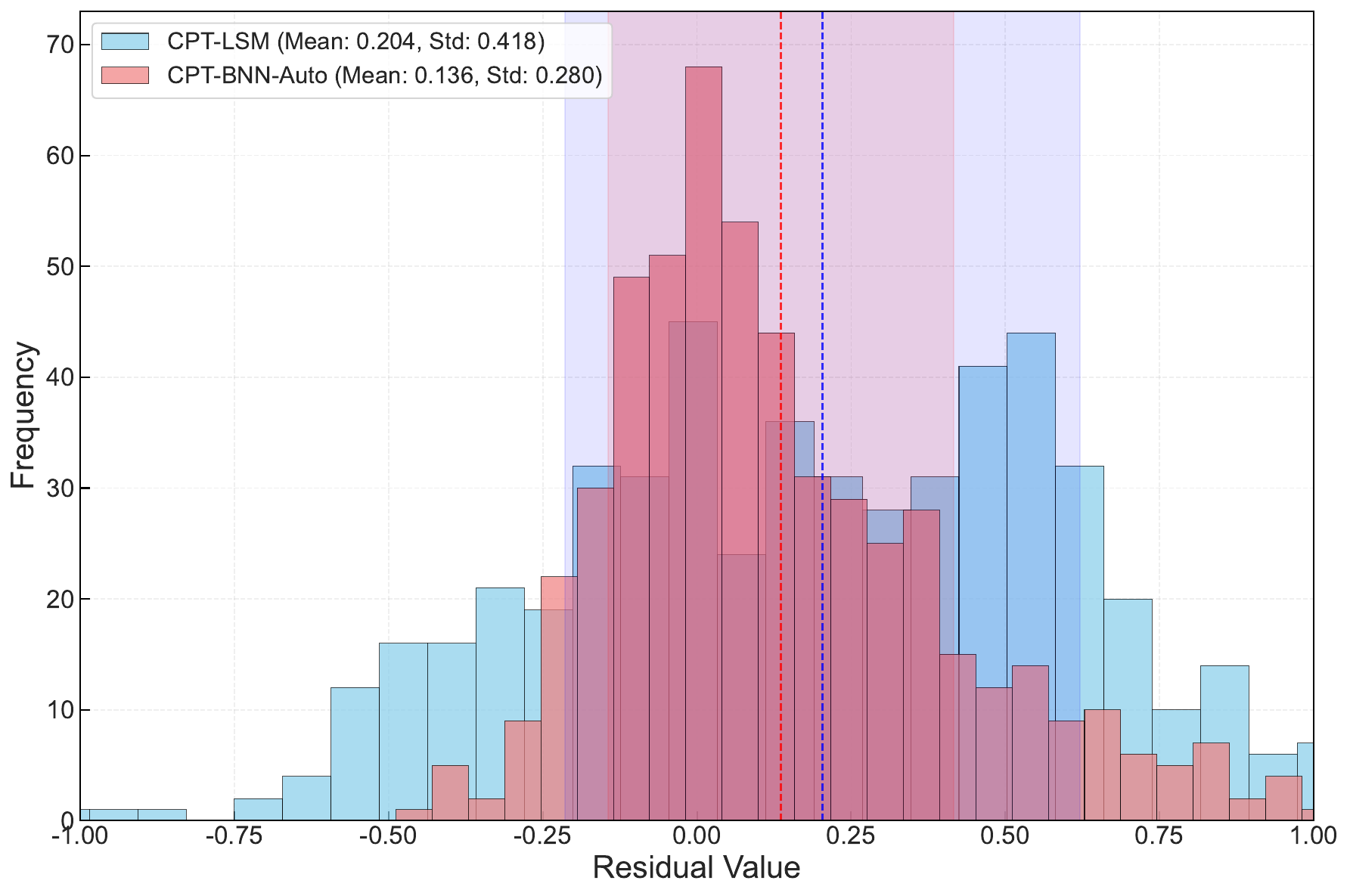}
 \caption{Comparison of the calculation of the mean and standard deviation for the logarithmic $\alpha$ decay half-lives using the CPT-LSM and CPT-BNN-Auto methods.}
 \label{Figure4}
 \end{figure}

To visually demonstrate the global improvement brought by the BNN-Auto method, Figs. \ref{Figure3} and \ref{Figure4} quantitatively compare the consistency between the calibrate and original calculations and the experimental ones from two dimensions of the spatial and statistical distribution. Fig. \ref{Figure3} presents a panoramic view of the spatial distribution of prediction residuals for the logarithmic half-lives of all 535 nuclides in the form of a nuclide chart ($N-Z$ diagram). The left CPT-LSM shows that the residuals of the conventional method are widely scattered across the whole range and exhibit regional and systematic clustering near specific neutron shells ($N=126$) or proton shells ($Z=82$). This clearly reveals inherent systematic biases in the theoretical model due to insufficient description of nuclear structure effects such as shell closures. In contrast, the scatter points in the right CPT-BNN-Auto are highly concentrated around the white region representing near-zero residuals. The colour distribution across the nuclide chart becomes uniform and neutral, and the distinct positive/negative bias colour blocks associated with shell closures that are apparent in the CPT-LSM plot have essentially disappeared. This strongly demonstrates that the BNN-Auto method, through data driven learning, effectively captures and corrects these systematic biases related to nuclear structure, achieving a uniform enhancement of prediction accuracy over the entire nuclide landscape. In addition, Fig. \ref{Figure4} further quantifies this global improvement by means of statistical histograms of the residuals. The residual distribution of CPT-LSM (blue) is broad with ranging approximately from –1.0 to over +1.0 and exhibits a flat, dispersed shape. In comparison, the residual distribution of CPT-BNN-Auto (pink) displays a sharp and tall peak, densely concentrated within the interval of –0.25 to +0.5, with the highest frequency occurring close to zero. Key statistical measures clearly corroborate this visual contrast, BNN-Auto improves the mean residual from 0.204 to 0.136, while significantly reducing the standard deviation from 0.418 to 0.280 with a reduction of $33\%$. The substantial decrease in standard deviation indicates that the dispersion and uncertainty of the predicted values have been effectively compressed. Together with the mean shifting closer to zero, these results confirm that BNN-Auto not only improves prediction accuracy but also greatly enhances precision and reliability. Consequently, the vast majority of predicted values now fall within 1.91 times the experimental error margin, providing a solid statistical foundation for robust extrapolative predictions in regions of superheavy nuclei where experimental data are scarce.

\begin{figure}[!htb]\centering
 \includegraphics
  [width=1.0\hsize]
  {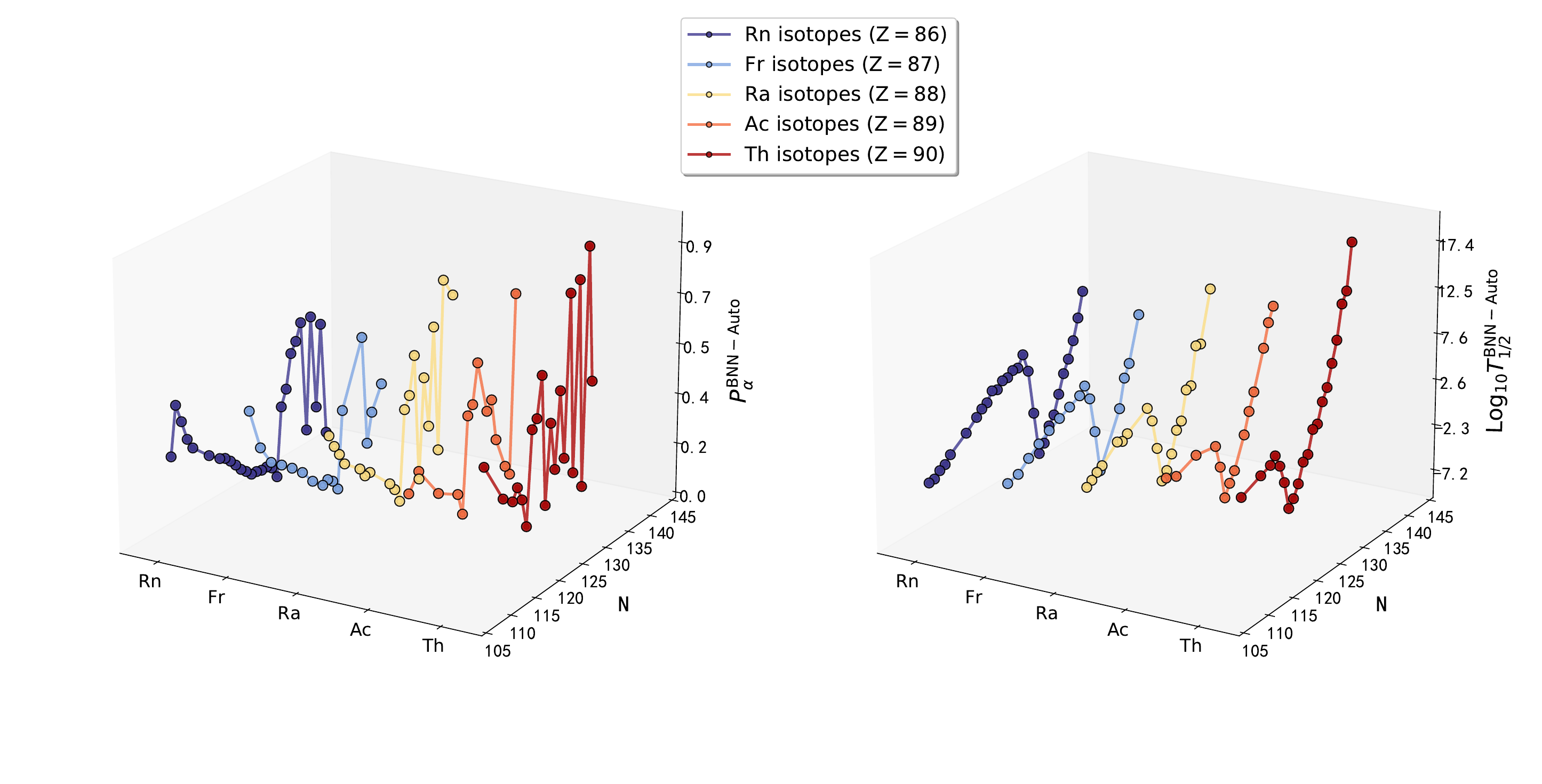}
 \caption{BNN-Auto calculated $\alpha$-particle preformation factors $P_{\alpha}^{\rm{BNN-Auto}}$ and half-lives $\rm{log}_{10}{T}_{1/2}^{\rm{BNN-Auto}}$ as a function of neutron number for isotopic chains $Z=86-90$.}
 \label{Figure5}
 \end{figure}

\begin{figure}[!htb]\centering
 \includegraphics
  [width=1.0\hsize]
  {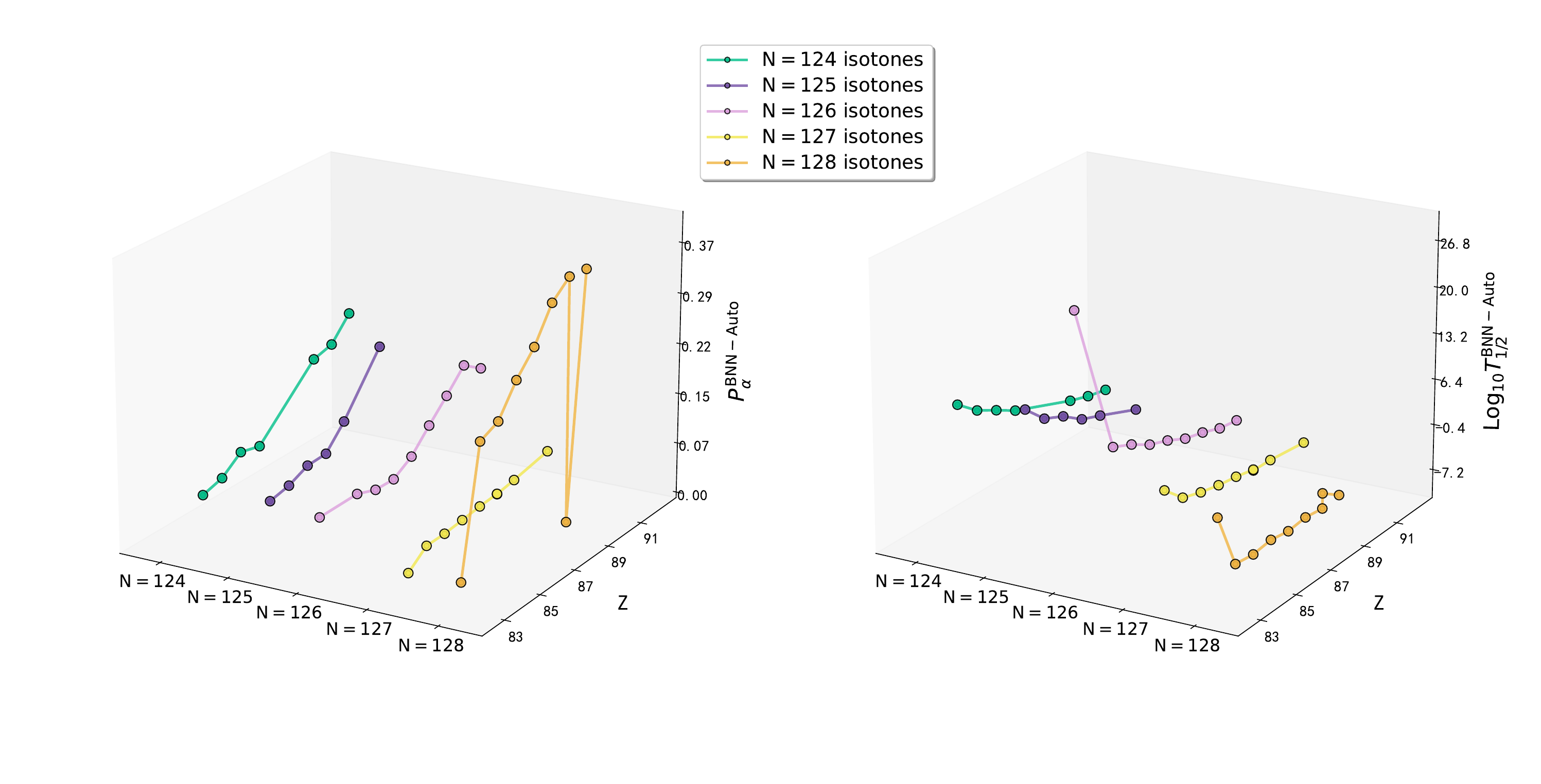}
 \caption{The $\alpha$-particle preformation factors $P_{\alpha}^{\rm{BNN-Auto}}$ and half-lives $\rm{log}_{10}{T}_{1/2}^{\rm{BNN-Auto}}$ calculated by the BNN-Auto method as a function of proton number for isotones with $N=124-128$.}
 \label{Figure6}
 \end{figure}

\begin{figure}[!htb]\centering
 \includegraphics
  [width=1.0\hsize]
  {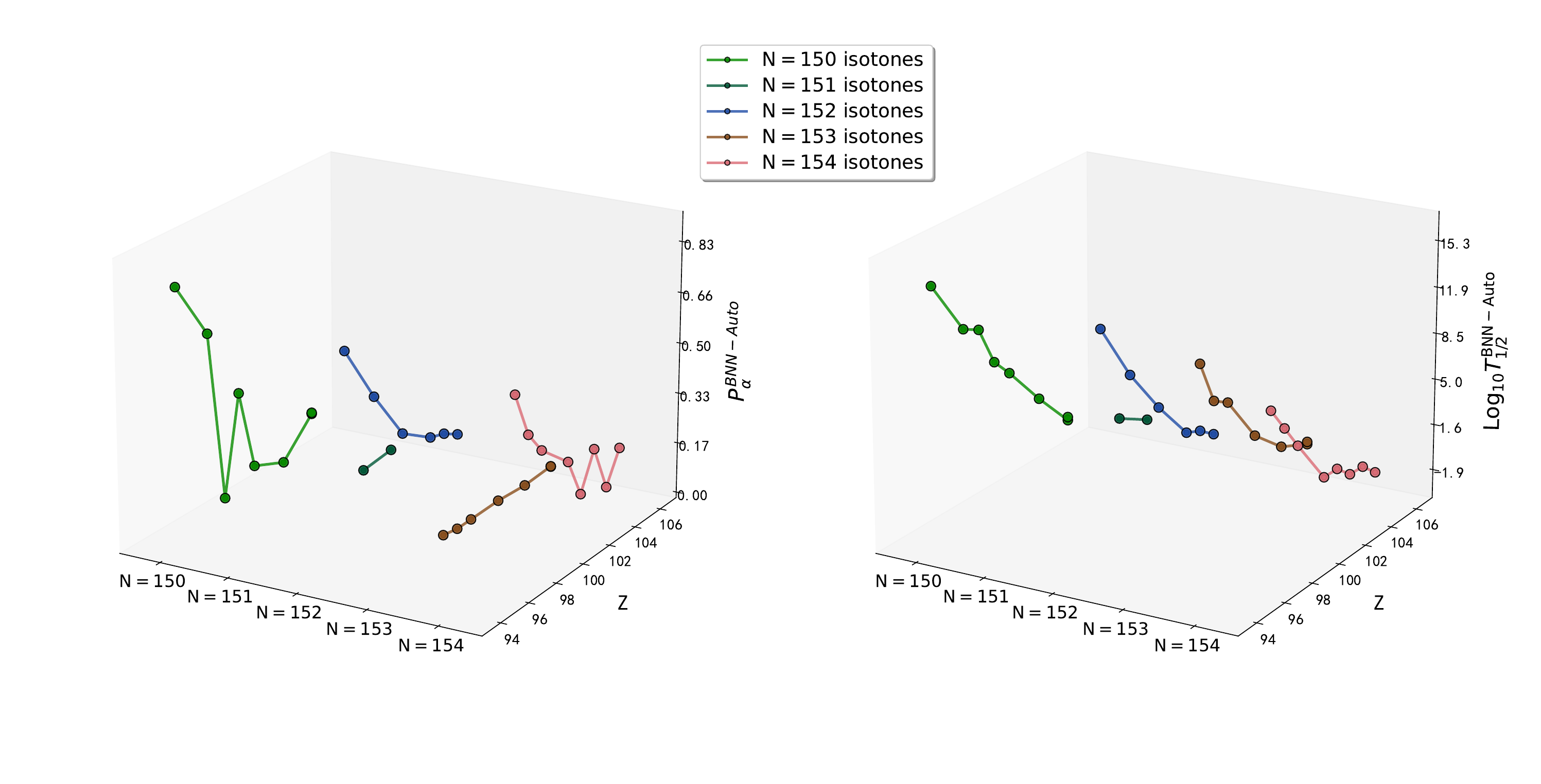}
 \caption{The $\alpha$-particle preformation factors $P_{\alpha}^{\rm{BNN-Auto}}$ and half-lives $\rm{log}_{10}{T}_{1/2}^{\rm{BNN-Auto}}$ calculated by the BNN-Auto method as a function of proton number for isotones with $N=150-154$.}
 \label{Figure7}
 \end{figure}

To further investigate the underlying physical mechanisms of the prediction results, we analyze the presence of odd-even effects in both the preformation factor $P_{\alpha}^{\mathrm{BNN-Auto}}$ and half-life ${T}_{1/2}^{\rm{BNN-Auto}}$. The odd-even staggering effect is a phenomenon widely observed in nuclear physics, manifesting in various quantities such as nuclear masses, single- and two-nucleon separation energies, decay energies, and half-life \cite{Satula:1998ha,Hove:2013fgd,Tang:2025uxy,Sun:2016jnb,Sun:2016bbw}. Whether this effect also influences the $\alpha$-particle preformation factor thus represents a pertinent question worthy of in-depth exploration. Figs. \ref{Figure5}-\ref{Figure7} present the nuclear structure information extracted from the decay half-lives and preformation factors along proton-neutron odd-even isotopic chains, as obtained via the BNN-Auto method.

Fig. \ref{Figure5} (left) shows the variation of the $\alpha$-particle preformation factors along the isotopic chains of $Z=$86-90 obtained by the BNN-Auto method with respect to the number of neutrons, while the right panel shows the corresponding logarithmic $\alpha$ decay half-lives, calculated within the CPT framework incorporating $P_{\alpha}^{\mathrm{BNN-Auto}}$, for the same isotopic chains. From this figure, a clear odd-even staggering is observed in both the preformation factors and the half-lives. The $P_ {\alpha}^{\mathrm{BNN-Auto}}$  for even-Z nuclei ($Z=$86, 88, 90) are generally higher than those for odd-Z nuclei ($Z=$87, 89), and the half-lives are relatively shorter. This reflects the higher stability of even-even nuclei due to pairing correlations, which enhances the probability of $\alpha$-particle preformation. At the same time, a pronounced peak in $P_{\alpha}^{\mathrm{BNN-Auto}}$ emerges near the magic neutron number $N=$126, particularly evident for nuclei close to $Z=$86, where the half-life decreases markedly. This indicates the dominant role of shell effects in this region.
For these isotopic chains with fixed $Z$, $N$ has a more significant influence on both $P_{\alpha}^{\mathrm{BNN-Auto}}$ and ${T}_{1/2}^{\rm{BNN-Auto}}$ than $Z$. Variations in the neutron number directly modulate the nuclear shell structure and pairing correlations, leading to the observed systematic trends.

Fig. \ref{Figure6} illustrates the variation of the $\alpha$-particle preformation factor with proton number $Z$ (left panel) and the logarithmic $\alpha$ decay half-lives of isotonic chains (right panel), within the CPT framework that incorporates $P_{\alpha}^{\mathrm{BNN-Auto}}$. The nuclei shown have neutron numbers $N=124-128$. From the figure, the odd-even alternation phenomenon is evident: nuclei with even $N$ exhibit higher $P_{\alpha}^{\mathrm{BNN-Auto}}$ values and correspondingly shorter ${T}_{1/2}^{\rm{BNN-Auto}}$, whereas the opposite trend holds for odd-N nuclei. As the proton number approaches the magic number $Z=82$ (near $N=126$), an anomalous variation occurs in $P_{\alpha}^{\mathrm{BNN-Auto}}$, accompanied by an increase in $\rm{log}_{10}{T}_{1/2}^{\rm{BNN-Auto}}$, indicating a suppression effect of the proton shell closure on the preformation factor. It is noteworthy that near $N=126$, the nuclides show lower $P_{\alpha}^{\mathrm{BNN-Auto}}$ and longer ${T}_{1/2}^{\rm{BNN-Auto}}$, consistent with shell effects in the heavy nuclear region.

The left panel of Fig. \ref{Figure7} presents the systematics of the $\alpha$-particle preformation factor as a function of $Z$ for nuclei sharing the same neutron numbers $N=150-154$. The corresponding $\rm{log}_{10}{T}_{1/2}^{\rm{BNN-Auto}}$ for these isotonic chains, calculated within the CPT framework using $P_{\alpha}^{\mathrm{BNN-Auto}}$, are displayed on the right panel. 
The odd-even staggering effect is evident across these chains as well, though it shows a more complex trend. In the superheavy region $Z>100$, the $P_{\alpha}^{\mathrm{BNN-Auto}}$ values are generally suppressed, resulting in characteristically longer half-lives that reflect the enhanced stability challenges of these nuclei. A local extremum in $P_{\alpha}^{\mathrm{BNN-Auto}}$, accompanied by a plateau in the half-life variation, is observed around $N=152$, suggesting the influence of a possible neutron subshell closure. 
This underscores the critical role of neutron effects on $\alpha$ preformation in the superheavy mass region. 
Compared to the trends in Fig. \ref{Figure6}, the behavior of these isotonic chains is dominated by neutron effects. This predominance stems from the fact that, in this high-N region, enhanced nucleonic correlations and shell effects outweigh the relatively weaker modulation from the varying proton Coulomb potential. These observations further corroborate that the neutron number is a primary driver of the $\alpha$ preformation factor, especially for nuclei far from the $\beta$-stability line. The BNN-Auto approach not only improves the predictive accuracy for $P_{\alpha}$ and ${T}_{1/2}$ but also, through visual analysis, elucidates key nuclear structure effects such as odd-even staggering and shell closures. These results highlight the critical importance of incorporating microscopic nuclear structure information into the theoretical description of $\alpha$ decay.

 \begin{figure}[!htb]\centering
 \includegraphics
  [width=0.7\hsize]
  {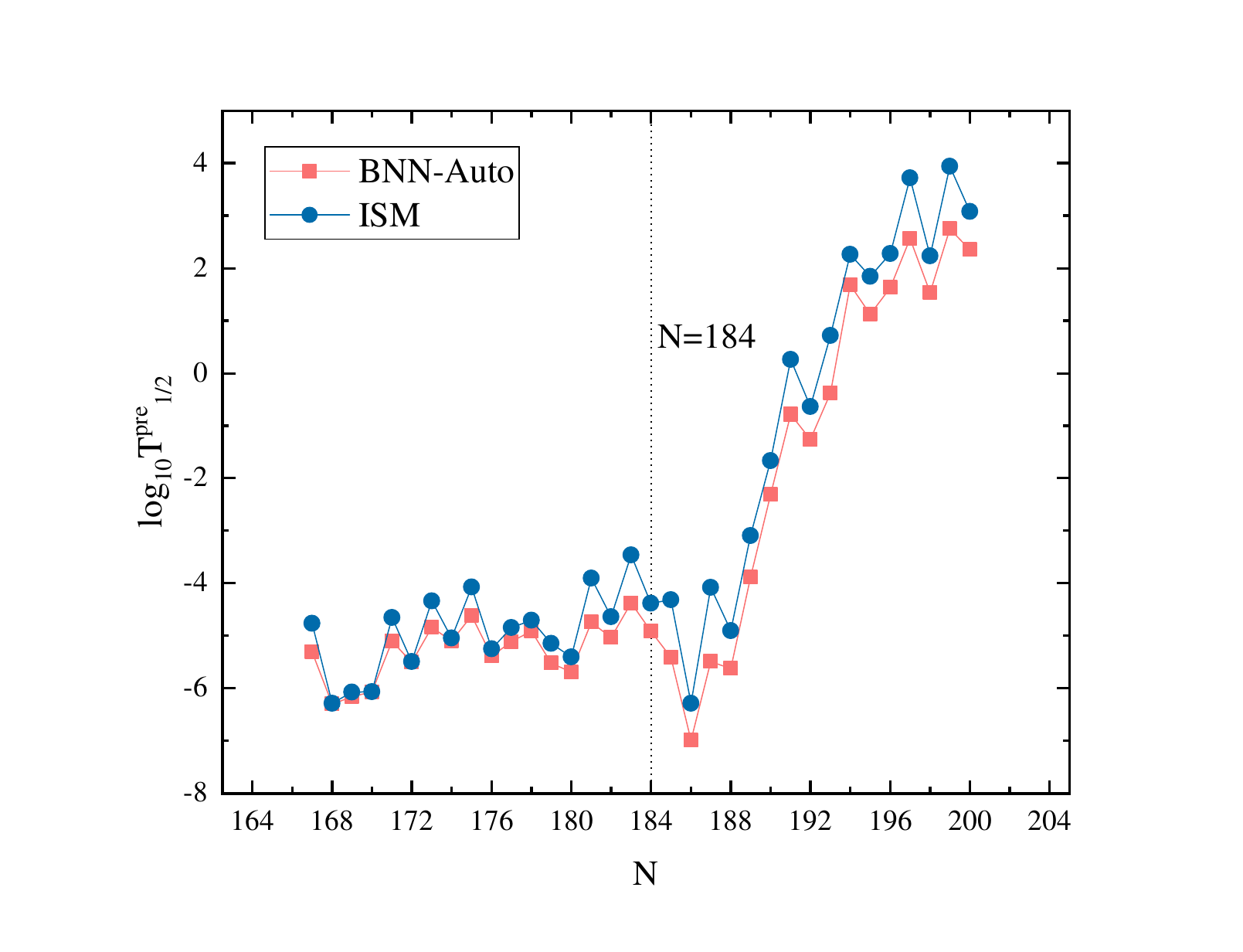}
 \caption{Predicted $\alpha$ decay half-lives in logarithmic form for $Z=120$ using the BNN-Auto and ISM, with $Q_{\alpha}$ taken from the WS4+ mass model \cite{Wang:2014qqa}.}
 \label{Figure8}
 \end{figure} 
To demonstrate the predictive capability of the present method, we further applied it to predict the corresponding half-lives of unknown superheavy nuclei with $Z=120$. The $Q_{\alpha}$ value plays a crucial role in predicting $\alpha$ decay half-lives, and the estimation of the corresponding preformation factor is also related to the decay energy. For a more scientific prediction, the reliable mass model WS4+ is adopted to predict the decay energies \cite{Wang:2014qqa}. The detailed results are listed in Table \ref{table5}. In this table, the first four columns present the $\alpha$ transition, the predicted $\alpha$ decay energy derived from the WS4+ mass model, $l$, and $P_{\alpha}^{\mathrm{BNN-Auto}}$, respectively. The fifth column gives the predicted $\alpha$ decay half-life calculated from $P_{\alpha}^{\mathrm{BNN-Auto}}$, denoted as $\rm{log}_{10}{T}_{1/2}^{\rm{Pre}}$. For reliability, the predictions from Ref. \cite{Zhu:2024swx} are taken as a reference. The table indicates that the half-lives predicted by BNN-Auto are generally consistent with those of the reference model, while exhibiting a more reasonable trend in the high-Z region $Z=120$. As shown in Fig. \ref{Figure8}, the predicted $\alpha$ decay half-lives from both the BNN-Auto and ISM models show a significant increase near N=184, a region associated with a possible island of stability. This common feature reflects the shell stabilization effect. This verifies the potential of BNN-Auto in predicting superheavy nuclei and provides theoretical guidance for future experiments.

\renewcommand\arraystretch{0.5}
\setlength{\tabcolsep}{3.6mm}
\begin{longtable}{cccccc}
	\caption{Predictions of $\alpha$ decay half-lives for superheavy nuclei with $Z=120$ using $Q_{\alpha}$ from the WS4+ mass model \cite{Wang:2014qqa}. }\\
	\hline
    \toprule
	{$\alpha$ transition}  & $Q_{\alpha}$  & $l$ &$P_{\alpha}^{\rm{BNN-Auto}}$ & $\rm{log}_{10}{T}_{1/2}^{\rm{Pre}}$& $\rm{log}_{10}{T}_{1/2}^{\rm{ISM}}$ \\
 	\hline
	\endfirsthead
	\multicolumn{1}{l}
	{Continued.} \\
\hline
    \toprule
	{$\alpha$ transition}  & $Q_{\alpha}$  & $l$ &$P_{\alpha}^{\rm{BNN-Auto}}$ & $\rm{log}_{10}{T}_{1/2}^{\rm{Pre}}$& $\rm{log}_{10}{T}_{1/2}^{\rm{ISM}}$ \\
	\hline
	\endhead
	\hline \\
	\endfoot
	\endlastfoot
$	^{287}	$120$	\to	^{283}	$Og$	$	&	13.91	&	4	&	0.0307 	&	-5.3089 	&	-4.7637 	\\
$	^{288}	$120$	\to	^{284}	$Og$	$	&	13.75	&	0	&	0.1344 	&	-6.2987 	&	-6.2834 	\\
$	^{289}	$120$	\to	^{285}	$Og$	$	&	13.68	&	0	&	0.1321 	&	-6.1628 	&	-6.0688 	\\
$	^{290}	$120$	\to	^{286}	$Og$	$	&	13.63	&	0	&	0.1307 	&	-6.0665 	&	-6.0663 	\\
$	^{291}	$120$	\to	^{287}	$Og$	$	&	13.41	&	2	&	0.0504 	&	-5.1027 	&	-4.6550 	\\
$	^{292}	$120$	\to	^{288}	$Og$	$	&	13.31	&	0	&	0.1232 	&	-5.5094 	&	-5.4938 	\\
$	^{293}	$120$	\to	^{289}	$Og$	$	&	13.24	&	2	&	0.0515 	&	-4.8354 	&	-4.3390 	\\
$	^{294}	$120$	\to	^{290}	$Og$	$	&	13.07	&	0	&	0.1237 	&	-5.0985 	&	-5.0454 	\\
$	^{295}	$120$	\to	^{291}	$Og$	$	&	13.1	&	2	&	0.0543 	&	-4.6179 	&	-4.0714 	\\
$	^{296}	$120$	\to	^{292}	$Og$	$	&	13.19	&	0	&	0.1341 	&	-5.3916 	&	-5.2517 	\\
$	^{297}	$120$	\to	^{293}	$Og$	$	&	13.02	&	0	&	0.1393 	&	-5.1067 	&	-4.8440 	\\
$	^{298}	$120$	\to	^{294}	$Og$	$	&	12.9	&	0	&	0.1472 	&	-4.9108 	&	-4.7026 	\\
$	^{299}	$120$	\to	^{295}	$Og$	$	&	13.19	&	0	&	0.1568 	&	-5.5183 	&	-5.1431 	\\
$	^{300}	$120$	\to	^{296}	$Og$	$	&	13.29	&	0	&	0.1673 	&	-5.6929 	&	-5.4064 	\\
$	^{301}	$120$	\to	^{297}	$Og$	$	&	13.03	&	2	&	0.0757 	&	-4.7377 	&	-3.9037 	\\
$	^{302}	$120$	\to	^{298}	$Og$	$	&	12.88	&	0	&	0.1922 	&	-5.0364 	&	-4.6392 	\\
$	^{303}	$120$	\to	^{299}	$Og$	$	&	12.8	&	2	&	0.0871 	&	-4.3730 	&	-3.4581 	\\
$	^{304}	$120$	\to	^{300}	$Og$	$	&	12.75	&	0	&	0.2212 	&	-4.9105 	&	-4.3786 	\\
$	^{305}	$120$	\to	^{301}	$Og$	$	&	13.27	&	2	&	0.1011 	&	-5.4074 	&	-4.3168 	\\
$	^{306}	$120$	\to	^{302}	$Og$	$	&	13.82	&	0	&	0.2390 	&	-6.9890 	&	-6.2875 	\\
$	^{307}	$120$	\to	^{303}	$Og$	$	&	13.58	&	4	&	0.0800 	&	-5.4919 	&	-4.0801 	\\
$	^{308}	$120$	\to	^{304}	$Og$	$	&	13.04	&	0	&	0.2791 	&	-5.6256 	&	-4.9014 	\\
$	^{309}	$120$	\to	^{305}	$Og$	$	&	12.17	&	0	&	0.2931 	&	-3.8840 	&	-3.0960 	\\
$	^{310}	$120$	\to	^{306}	$Og$	$	&	11.48	&	0	&	0.2927 	&	-2.3158 	&	-1.6677 	\\
$	^{311}	$120$	\to	^{307}	$Og$	$	&	11.1	&	2	&	0.1137 	&	-0.7835 	&	0.2607 	\\
$	^{312}	$120$	\to	^{308}	$Og$	$	&	11.05	&	0	&	0.2900 	&	-1.2608 	&	-0.6333 	\\
$	^{313}	$120$	\to	^{309}	$Og$	$	&	10.92	&	2	&	0.1178 	&	-0.3757 	&	0.7179 	\\
$	^{314}	$120$	\to	^{310}	$Og$	$	&	9.97	&	0	&	0.3065 	&	1.6794 	&	2.2649 	\\
$	^{315}	$120$	\to	^{311}	$Og$	$	&	10.15	&	0	&	0.3002 	&	1.1229 	&	1.8448 	\\
$	^{316}	$120$	\to	^{312}	$Og$	$	&	9.97	&	0	&	0.3123 	&	1.6291 	&	2.2780 	\\
$	^{317}	$120$	\to	^{313}	$Og$	$	&	9.96	&	3	&	0.0956 	&	2.5627 	&	3.7229 	\\
$	^{318}	$120$	\to	^{314}	$Og$	$	&	9.99	&	0	&	0.3149 	&	1.5381 	&	2.2329 	\\
$	^{319}	$120$	\to	^{315}	$Og$	$	&	9.89	&	3	&	0.0911 	&	2.7453 	&	3.9411 	\\
$	^{320}	$120$	\to	^{316}	$Og$	$	&	9.71	&	0	&	0.3493 	&	2.3543 	&	3.0772 	\\
\toprule	
\label{table5}
\end{longtable}

\section{Summary}
\label{Sec.IV}
In this work, a novel framework for describing the $P_{\alpha}$ is constructed by integrating the BNN-Auto. This framework quantifies the uncertainty of model parameters based on Bayesian probability, overcoming the limitations of traditional least squares method in parameter optimization. The results show that BNN-Auto can significantly improve the prediction accuracy of $P_{\alpha}$ and half-life, and demonstrates excellent generalization ability in both even-even and odd-odd nuclei. By analyzing the variation patterns of $P_{\alpha}$ in different isotopic chains and isotones, this study reveals the inhibitory effect of neutron shell closure on the preformation probability, as well as the blocking effect of unpaired nucleons in odd-even nuclei on $P_{\alpha}$. Furthermore, predictions for superheavy nuclei with Z=120 align with existing theoretical models and show more physically reasonable trends in half-lives. The BNN-Auto framework not only provides a highly reliable theoretical tool for the study of $\alpha$ decay, but also offers a new data-driven paradigm for understanding the evolution of nuclear structure under extreme conditions. It provides certain valuable guidance for future predictions in the superheavy nucleus region.

\begin{acknowledgments}
Supported by the National Natural Science Foundation of China (Grants Nos: 12175100, 11975132 and 12405154).
\end{acknowledgments}

\bibliography{ref}

\end{document}